\documentclass[twocolumn,aps,showpacs]{revtex4}
\usepackage{latexsym}
\usepackage[dvips]{graphics}
\usepackage{amssymb}

\begin{document}

\title{Evolution of a Vacuum Shell in the Friedman-Schwarzschild
World}

\author{V. I. Dokuchaev}
 \email{dokuchaev@ms2.inr.ac.ru}
 \affiliation{Institute for Nuclear Research of the Russian
 Academy of Sciences, 60-letiya Oktyabrya pr. 7a, Moscow, 117312
 Russia}
\author{S. V. Chernov}
 \email{chernov@td.lpi.ru}
 \affiliation{Lebedev Physical Institute of the Russian Academy of
 Sciences, Leninskii pr. 53, Moscow, 119991 Russia}
 \affiliation{Institute for Nuclear Research of the Russian
 Academy of Sciences, 60-letiya Oktyabrya pr. 7a, Moscow, 117312
 Russia}

\date{\today}

\begin{abstract}
The method of an effective potential is used to investigate the
possible types of evolution of vacuum shells in the Friedman-
Schwarzschild world. Such shells are assumed to emerge during
phase transitions in the early Universe. The possible global
geometries are constructed for the Friedman- Schwarzschild worlds.
Approximate solutions to the equation of motion of a vacuum shell
have been found. The conditions under which the end result of the
evolution of the vacuum shells under consideration is the
formation of black holes and wormholes with baby universes inside
have been found. The interior of this world can be a closed, flat,
or open Friedman universe.
\end{abstract}
 \maketitle

\section{INTRODUCTION}

Phase transitions in the early Universe occur with the formation
of vacuum bubbles of a new phase. During the expansion and mutual
intersection of new-phase bubbles, old-phase bubbles completely
surrounded by the new phase can also be formed
\cite{Kirznits,Kirznits2,ZeldovKobzarOcyn,KobzOkunVoloshin,Col,
CalCol,ColLuc}. Investigating the evolution of such bubbles and
their subsequent fate is of interest in connection with the
problems of primordial black holes. The evolution of vacuum
bubbles was considered in many papers mainly under the assumption
of a de Sitter metric for the bubble interior. The region that
separates the bubble interior and exterior is a domain wall. The
shin-shell formalism suggested by Israel \cite{Israel} and
subsequently developed in detail by Berezin, Kuzmin, and Tkachev
\cite{BerKuzTkach}, as applied to cosmological problems, is
commonly used to describe the latter. Various special cases of
this problem were considered in many papers on cosmological phase
transitions (see, e.g., the early papers
\cite{ZeldovKobzarOcyn,KobzOkunVoloshin,Col,ColLuc,Sato,
SatoSasKodMae1,Sato1,Sato2}). The end result of the evolution of
vacuum bubbles can be the formation of primordial black holes and
various types of wormholes with baby universes inside
\cite{Sato,BerKuzTkach,BerKuzTkach1,IpserSikivie,Aurilia,Aurilia1,
Aurilia2,Aguirre,BlGuenGuth,BerKuzTkach2,BerKuzTkach6,BerKuzTkach4,
BerKuzTkach5,Kardashev}. The evolution of vacuum bubbles in the
Schwarzschild-de Sitter world was investigated in
\cite{BlGuenGuth,BerKuzTkach,DokCher,DokCher1}. The dynamics of a
bubble in the Friedman-Schwarzschild world was investigated in
\cite{Sato,SatoSasKodMae1} but without including the surface
tension of the bubble (shell) wall. In this paper, we analyze the
full dynamics of a vacuum shell in the Friedman- Schwarzschild
world in the thin-wall approximation by taking into account the
surface energy density of the shell. Such a configuration can
result from the production of particles during the vacuum decay
inside the bubble by analogy with the final stage of inflation
(see also \cite{Rubakov}). As a result of our analysis, we found
all of the possible types of evolution of vacuum shells in the
Friedman-Schwarzschild world and constructed the corresponding
global geometries. We also found approximate asymptotic solutions
to the equation of motion of vacuum shells in the Friedman-
Schwarzschild world.

\section{A SHELL IN THE FRIEDMAN- SCHWARZSCHILD WORLD}

Let us consider a spherically symmetric shell whose interior (far
from the boundary) is described by the Friedman metric
\begin{equation}
 ds^2=dt_{\rm in}^2-a^2(t_{\rm in})\left[\frac{dq^2}
 {1-kq^2}+q^2d\Omega^2\right],
\end{equation}
where $t_{\rm in}$ is the time of an observer in the Friedman
world, $a=a(t_{\rm in})$ is the scale factor, $q$ is the inner
radial coordinate, $d\Omega$ is an element of the solid angle,
$k=1$, $k=0$, and $k=-1$ for closed, flat, and open worlds,
respectively. The exterior of the shell is described by the
Schwarzschild metric
\begin{equation}
 ds^2=\left(1-\frac{2m}{r}\right)dt_{\rm out}^2-
 \left(1-\frac{2m}{r}\right)^{-1}dr^2-r^2d\Omega^2,
\end{equation}
where m is the total outer mass of the shell and r is the outer
radial coordinate. In what follows, the subscripts "in" and "out"
pertain to the inner Friedman and outer Schwarzschild worlds,
respectively. The metric of the transition region is modelled in
the form of a thin shell $\Sigma$:
\begin{equation}
 ds^2|_{\Sigma}=d\tau^2-\rho^2(\tau)d\Omega^2,
\end{equation}
where $\tau$ is the proper time of an observer on the shell and
$\rho=\rho(\tau)$ is the shell radius. The inner and outer metrics
are joined on the shell using the thin-shell method \cite{Israel}
to give the equations of motion of the shell \cite{BerKuzTkach}
\begin{equation}
 4\pi S_{0}^{0}=[K_{2}^{2}], \quad
 \frac{dS_{0}^{0}}{d\tau}+2(S_{0}^{0}-
 S_{2}^{2})\frac{\dot{\rho}}{\rho}+[T_{0}^{n}]=0,
 \label{eqmotion}
\end{equation}
where $S_{\alpha}^{\beta}$ is the surface energy density tensor on
the shell, $K_{\alpha}^{\beta}$ is the external curvature tensor,
$T_{\alpha}^{\beta}$ is the fluid energy-momentum tensor, and $[A]
= A_{out} - A_{in}$. For a vacuum shell,
$S_{0}^{0}=S_{2}^{2}=S=const$. Let us first consider the interior
of our bubble. The Friedman equations are
\begin{equation}
 \frac{\dot{a}^2+k}{a^2}=\frac{8\pi}{3}\varepsilon, \qquad
 \frac{\ddot{a}}{a}=-\frac{4\pi}{3}(\varepsilon+3P),
 \label{friedman}
\end{equation}
where the dot denotes differentiation with respect to the time
$t_{in}$, $\varepsilon$ is the energy density, and $P$ is the
pressure. We will make the classification for an arbitrary
equation of state, but we will always keep in mind a linear
equation of state, where $P=\alpha\varepsilon$ and
$\alpha=const\neq-1$ (for the classification of solutions in the
case of $\alpha=-1$ corresponding to the de Sitter vacuum metric,
see \cite{DokCher,DokCher1}). For a linear equation of state, the
solution to the Friedman equations is ($k = 0$)
\begin{equation}
 a=At_{in}^n, \quad \varepsilon=\frac{3n^2}{8\pi t_{in}^2}, \quad
 \label{frsolution}
\end{equation}
where $A$ is a constant and $n=2/(3(1+\alpha))$. For the Friedman
metric and using the condition for joining the Friedman metric and
the shell, $\rho=aq$, Berezin et al. \cite{BerKuzTkach} calculated
(for any k) the invariants
\begin{equation}
 \Delta\equiv g^{\alpha\beta}\rho_{,\alpha}\rho_{,\beta}=
 \frac{8\pi}{3}\varepsilon\rho^2-1
 \label{Delta}
\end{equation}
and the external curvature tensor component (since the problem is
spherically symmetric, we will need only one component)
\begin{equation}
 K_{2}^{2}=
 -\frac{\sigma}{\rho}\sqrt{\left(\frac{d\rho}{d\tau}\right)^2
 +1-\frac{8\pi}{3}\varepsilon\rho^2},
\end{equation}
where  $\sigma=\pm1$; $\sigma=1$ if the radius of the
two-dimensional sphere increases in the direction of the outward
normal and $\sigma=-1$ in the opposite case. In turn, depending on
the sign of the invariant $\Delta$, the shell moves either in the
space-time region $R$ or in $T$ \cite{BerKuzTkach3}. The boundary
that separates the space-time regions $R$ and $T$ for the Friedman
metric is located at the radius
\begin{equation}
 \rho_{\Delta}=\sqrt{\frac{3}{8\pi\varepsilon}},
 \label{rhodelta}
\end{equation}
which is the root of the equation $\Delta(\rho)=0$. For the outer
Schwarzschild metric, the external curvature tensor component
$K_{2}^{2}$ is
\begin{equation}
 K_{2}^{2}=
 -\frac{\sigma}{\rho}\sqrt{\left(\frac{d\rho}{d\tau}\right)^2+1
 -\frac{2m}{\rho}}\,,
\end{equation}
and the radius at which $\Delta$ changes its sign coincides with
the radius of the event horizon $r_h=2m$. As a result, the main
equation of motion of the shell (\ref{eqmotion}) that arises as a
condition for joining the outer and inner metrics can be written
for any $k$ as \cite{BerKuzTkach}
\begin{equation}
 4\pi S=
 \frac{\sigma_{\rm in}}{\rho}\sqrt{\left(\frac{d\rho}{d\tau}\right)^2
 +1-\frac{8\pi}{3}\,\varepsilon\rho^2}
 -\frac{\sigma_{\rm out}}{\rho}\sqrt{\left(\frac{d\rho}{d\tau}\right)^2
 +1-\frac{2m}{\rho}}\,.
 \label{eqmot}
\end{equation}\
In this equation, the radius $\rho$ depends on the proper time
$\tau$ of an observer on the shell and the energy density
$\varepsilon$ depends on the time $t_{\rm in}$ for an observer
inside the shell in the Friedman world. Therefore, the equation of
motion should be supplemented with another equation obtained when
the inner Friedman metric and the metric on the shell are joined:
\begin{equation}
 dt_{in}^2-a^2dq^2=d\tau^2.
 \label{shifka}
\end{equation}
In the Eq.~(\ref{eqmot}), the shell radius $\rho$ may be
considered as a function of the time $t_{\rm in}$ (below, we omit
the subscript to save space). More specifically, $\tau=\tau(t)$
can be expressed from Eq.~(\ref{shifka}) and substituted into Eq.~
(\ref{eqmot}), i.~e., $\rho(\tau)=\rho(\tau(t))$. We will begin
our analysis with the full classification of the solutions to the
equation of motion of the shell (\ref{eqmot}) and the construction
of the corresponding global geometries. Subsequently, we will find
an approximate solution to the equation of motion of the shell in
some special cases.

\section{ANALYSIS OF THE EQUATION OF MOTION OF THE SHELL}

For the subsequent analysis, it is convenient to represent the
equation of motion of the shell (\ref{eqmot}) as an equation for
the effective energy,
\begin{eqnarray}
\left(d\rho/d\tau\right)^2/2+U(\rho)=0 \nonumber
\end{eqnarray}
(see also \cite{DokCher,DokCher1}), where the effective potential
is
\begin{widetext}
\begin{equation}
 U(\rho)=\frac{1}{2}\Bigg[1-\left(2\pi S+
 \frac{\varepsilon}{3S}\right)^2\rho^2
 -\frac{m}{\rho}\left(1-\frac{\varepsilon}{6\pi S^2}\right)-
 \frac{m^2}{16\pi^2 S^2\rho^4}\Bigg].
 \label{potential}
\end{equation}
\end{widetext}
Its graph is presented in Fig.~1. Equation (\ref{potential}) for
the effective potential $U(\rho)$ should be supplemented with the
following conditions on the signs of the quantity $\sigma$ present
in the original equation of motion of the shell (\ref{eqmot}):
\begin{eqnarray}
 \label{sigmain}
 \sigma_{\rm in}&=&
 sign\left[m-\frac{4\pi}{3}\varepsilon\rho^3+8\pi^2
 S^2\rho^3\right],
 \\
 \sigma_{\rm out}&=&sign\left[m-\frac{4\pi}{3}\varepsilon\rho^3
 -8\pi^2 S^2\rho^3\right].
 \label{sigmaout}
\end{eqnarray}
It is easy to show that the second derivative of this potential
with respect to the radius for any $\rho=\rho(\tau(t))$ is
negative:
\begin{widetext}
\begin{equation}
 \frac{\partial^2U}{\partial\rho^2}=
 -\frac{1}{2}\Bigg[\frac{2m}{\rho^3}+
 \frac{m^2}{\pi^2S^2\rho^6}+8\pi^2S^2+
 \frac{8\pi}{3}\varepsilon+\frac{\varepsilon^2}{9S^2}
 +\left(\frac{\varepsilon}{3S}-
 \frac{m}{2\pi S\rho^3}\right)^2\Bigg]<0\,.
\end{equation}
\end{widetext}
Thus, there are no static solutions in this problem
\cite{IshakLake}. Setting the first derivative of the potential
with respect to the radius equal to zero, we find the point of
maximum potential $\rho_{\rm max}^3=my_{\rm max}$, where
\begin{widetext}
\begin{equation}
 y_{\rm max}=\left[1-\frac{\varepsilon}{6\pi S^2}+
 \sqrt{\left(1-\frac{\varepsilon}{6\pi S^2}\right)^2+
 8\left(1+\frac{\varepsilon}{6\pi S^2}\right)^2}\,\right]
 \left(4\pi S+\frac{2\varepsilon}{3S}\right)^{-2}>0.
 \label{ymax}
\end{equation}
\end{widetext}
Note, that the point of maximum potential is a function of time,
$\rho_{\rm max}=\rho_{\rm max}(t)$. As will be shown below, the
total mass m (Schwarzschild mass) of the shell measured by an
observer at spatial infinity is a convenient parameter for the
classification of the possible types of its evolution. This mass
with the gravitational mass defect includes the total energy of
the inner Friedman world and the total energy of the shell with
its surface tension energy and its kinetic energy. Substituting
$\rho=\rho_{\rm max}$ into Eq.~(\ref{potential}) for the
potential, we find the first important mass parameter of the
shell, $m_{\rm max}=m_{0}$, at which a contracting or expanding
shell passes through the point of maximum potential:
\begin{widetext}
\begin{equation}
 m_{0}=\sqrt{y_{\rm max}}\Bigg[1-\frac{\varepsilon}{6\pi S^2}+
 \frac{1}{16\pi^2S^2y_{\rm max}}+\left(2\pi S+
 \frac{\varepsilon}{3S}\right)^2y_{\rm max}\Bigg]^{-3/2}>0.
 \label{m0}
\end{equation}
\end{widetext}
The potential $U(\rho_{\rm max})<0$ for $m>m_{0}$ and, conversely,
$U(\rho_{\rm max})>0$ for $m<m_{0}$. Thus, depending on the total
mass m of the shell, the potential either intersects the $U = 0$
axis or does not. In other words, this means that the presence or
absence of a bounce point during the temporal evolution of the
shell radius depends on the total mass m of the shell. It should
be kept in mind that the mass parameter $m_{0}$  (and all of the
mass parameters introduced below) is a function of time t, because
the energy density $\varepsilon$ depends on t in accordance with
the Friedman equations (\ref{friedman}). Therefore, at a fixed
total mass m, inequalities of the form $m>m_{0}$  or $m<m_{0}$ can
change with time to the opposite ones. Accordingly, the bounce
point can appear and/or disappear as the shell evolves. Note also
that the energy density $\varepsilon$  decreases with time for a
linear equation of state, $P=\alpha\varepsilon$, at $\alpha>-1$
and $k=0$. Therefore, on fairly long time scale, we have the
following asymptotic for the mass parameter:
$m_{0}(t\rightarrow\infty)=4/(27\pi S)$ Accordingly, the
inequality $U(\rho_{\rm max})\lessgtr0$ will hold on fairly long
time scales for $m\gtrless m_{0}(t\rightarrow\infty)$. Next, it
follows from Eq.~(\ref{sigmain}) that $\sigma_{\rm in}$ changes
its sign ($\sigma_{\rm in}=0$) at the shell radius
$\rho=\rho_{1}$, where
\begin{equation}
 \rho_{1}^3= \frac{3}{4\pi}\,\frac{m}{\varepsilon-6\pi S^2},
 \label{rho1}
\end{equation}
The radius $\rho_{1}$ exists only at $\varepsilon(t,k)>6\pi S^2$.
For a linear equation of state, $t<t_{1}=n/(4\pi S)$ at $k=0$. The
radius $\rho_{1}$ does not exist ($\rho_{1}<0$) at $t>t_{1}$. We
see from these relations that on time scales $t<t_{1}$ there
exists a radius in a flat universe at which $\sigma_{in}$ changes
its sign; the latter, in turn, is related to the regions $R_+$
(where $dr/dq > 0$) and $R_-$ (where $dr/dq < 0$). The solution of
the Friedman equations determines the time scales at which the
radius $\rho_{1}$ will exist for other equations of state. Note
also that a periodic function can be the solution of the Friedman
equations for a closed universe. Therefore, the radius $\rho_{1}$
can appear and disappear an infinite number of times. We will
assume that the radius appears only once. This is a very rough
approximation that can subsequently lead to contradiction on the
Carter-Penrose diagrams if this condition is disregarded. In turn,
it follows from Eq.~(\ref{sigmaout}) that $\sigma_{\rm out}$
changes its sign ($\sigma_{\rm out}=0$) at the shell radius
$\rho=\rho_{2}$, where
\begin{equation}
 \rho_{2}^3= \frac{3}{4\pi}\,\frac{m}{\varepsilon+6\pi S^2},
 \label{rho2}
\end{equation}
The relations between $\rho_{1}$, $\rho_{2}$ and $\rho_{\rm max}$
follow from Eqs. (\ref{ymax}), (\ref{rho1}) and (\ref{rho2}):
\begin{eqnarray}
 \rho_{2}&<&\rho_{\rm max}; \\
 \rho_{1}&>&(\rho_{2},\rho_{\rm max}) \quad \hbox{for}
 \quad \varepsilon>6\pi S^2.
\end{eqnarray}
Substituting the radius $\rho=\rho_{1}$  into
Eq.~(\ref{potential}) for the potential $U(\rho)$ and solving the
equation
\begin{equation}
 U(\rho_{1})\equiv\frac{1}{2}\left[1-
 2\!\left(\frac{4\pi}{3}\right)^{1/3}\!\!
 \left(\frac{m\varepsilon^{3/2}}{\varepsilon-
 6\pi S^2}\right)^{2/3}\right]=0.
\end{equation}
we find the mass parameter $m=m_1$, where
\begin{equation}
 m_{1}=\frac{1}{4}\sqrt{\frac{3}{2\pi}}
 \frac{\varepsilon-6\pi S^2}{\varepsilon^{3/2}}.
\end{equation}
We see from this relation that $U(\rho_{1})\lessgtr0$ for
$m\gtrless m_{1}$. This parameter exists, just as $\rho_{1}$, only
at $\varepsilon>6\pi S^2$. Similarly, substituting the radius
$\rho=\rho_{2}$ into Eq.~(\ref{potential}) for the potential
$U(\rho)$  and solving the equation
\begin{equation}
 U(\rho_{2})\!\equiv\!\frac{1}{2}\left[1\!-
 \!2\!\left(\frac{4\pi}{3}\right)^{1/3}\!\!\!m^{2/3}
 (\varepsilon\!+\!6\pi S^2)^{1/3}\right]=0,
\end{equation}
we find another mass parameter, $m=m_2$, where
\begin{equation}
 m_{2}=
 \frac{1}{4}\sqrt{\frac{3}{2\pi(\varepsilon+6\pi S^2)}}.
 \label{m2}
\end{equation}
According to Eq.~(\ref{frsolution}), the energy density in the
Friedman world decreases as $\varepsilon\propto t^{-2}$  (for a
linear equation of state and at $k=0$). Therefore, the mass
parameter $m_{2}$ increases with time:
\begin{equation}
 d m_{2}/dt>0, m_{2}(t\rightarrow\infty)\rightarrow(8\pi S)^{-1}
\end{equation}
For $m<m_{2}$, the potential is always positive at the point with
$\rho=\rho_{2}$, i.~e., $U(\rho_{2})>0$. In contrast, for
$m>m_{2}$, the potential is always negative at the point with
$\rho=\rho_{2}$, i.~e., $U(\rho_{2})<0$. In other words, the point
with coordinates $(\rho_{2},0)$ lies under the graph of $U(\rho)$
for $m<m_{2}$ and above the graph of $U(\rho)$ for $m>m_{2}$. At
$m=m_{2}$, the radius $\rho_{2}$ intersects the potential. Note
also that $m_{2}>m_{1}$. For the potential on the event horizon of
the Schwarzschild metric, $\rho_{h}=2m$, we find
\begin{equation}
 U(\rho_h)=-\left[\pi Sm\left(1+
 \frac{\varepsilon}{6\pi S^2}\right)-\frac{1}{64\pi Sm}\right]^2\leq0.
\end{equation}
We see that the point with coordinates $(\rho_h,0)$ always lies
either above the graph of $U(\rho)$ or touches the graph of the
potential at $m=m_{2}$ ($U(\rho_{h})=0$). At $m=m_2$, the radius
of the event horizon $\rho_{h}=2m$ coincides with the radius
$\rho_{2}$ at which $\sigma_{\rm out}$ changes its sign. Using
Eqs. (\ref{m0}), (\ref{rho2}) and (\ref{m2}) for $m_{0}$,
$\rho_{2}$ and $m_{2}$, we find that $m_{0}>m_{2}$ and
\begin{equation}
\rho_{h} \gtrless \rho_{2} \quad \hbox{for} \quad m \gtrless
m_{2}.
\end{equation}
Finally, substituting the radius of the boundary between the
space-time regions $R$ and $T$ in the Friedman world,
$\rho=\rho_{\Delta}$ from (\ref{rhodelta}), into
Eq.~(\ref{potential}) for the potential $U(\rho)$ yields
\begin{widetext}
\begin{equation}
 U(\rho_{\Delta})=-\frac{1}{2\rho_{\Delta}}\left[2\pi S\left(1-
 \frac{\varepsilon}{6\pi S^2}\right)
 \left(\frac{3}{8\pi\varepsilon}\right)^{3/4}+\frac{m}{4\pi S}
 \left(\frac{8\pi\varepsilon}{3}\right)^{3/4}\right]^2
 \leq0.
\end{equation}
\end{widetext}
We see that the point with coordinates $(\rho_{\Delta},0)$ cannot
be under the graph $U(\rho)$. Only at $m=m_{1}$ does the point
with coordinates $(\rho_{\Delta},0)$ lie on the graph of the
potential and, in this case, $\rho_{\Delta}=\rho_{1}$.
Accordingly, $\rho_{1}\gtrless\rho_{\Delta}$ for $m\gtrless
m_{1}$. Note that the radius $\rho_{\Delta}$ at which the regions
$R$ and $T$ are interchanged has the same properties as the radius
of the event horizon in the Schwarzschild metric, $r_h=2m$, with
respect to our potential. It can also be shown that
$\rho_{\Delta}>\rho_{\rm max}$ at $m=m_{\rm max}$. Finally, let us
introduce the last mass parameter
\begin{equation}
 m_{3}=\frac{1}{4}\sqrt{\frac{3}{2\pi\varepsilon}},
\end{equation}
which is the root of the equation $\rho_{\Delta}=\rho_{h}$. As a
result, we will obtain the relations
$\rho_{h}\lessgtr\rho_{\Delta}$ for $m\lessgtr m_{3}$. It can be
shown that $m_{3}>m_{0}$. For a linear equation of state,
$m_{3}(t\rightarrow\infty)\rightarrow\infty$ at $k = 0$. We now
have all of the necessary parameters to construct a full
classification of the possible types of solutions to the equation
of motion of the shell in the Friedman-Schwarzschild world and to
find the corresponding global geometries.

\section{GLOBAL GEOMETRIES OF THE FRIEDMAN-SCHWARZSCHILD WORLD}

Let us consider all of the possible types of solutions to the
equation of motion of a vacuum shell in the Friedman-Schwarzschild
world and then give a physical interpretation of these solutions.

\subsection{The Case of $m>m_{3}$}

We will begin our consideration with the case where $m>m_{3}$,
i.~e., where the shell has a large mass that exceeds all of the
characteristic masses in our problem, and will sequentially
consider shells with an increasingly small mass. In this case, the
relations $\rho_{h}>\rho_{\Delta}$, $\rho_{1}>\rho_{2}$ (if
$\rho_{1}$ exists), $\rho_{1}>\rho_{\Delta}$ and
$\rho_{h}>\rho_{2}$ hold. The potential and the location of the
characteristic radii for this case are shown in Fig.~2a. As we see
from Fig. 2a, there is no bounce point for the vacuum shell in
this case. Let us consider the special case where the vacuum shell
initially expands. To determine the type of space-time regions $R$
and $T$, it is important to know which signs $\sigma_{\rm in}$ and
$\sigma_{\rm out}$ will have when the shell intersects the radii
$\rho_{\Delta}$ and $\rho_{h}$, respectively. When the radius
$\rho_{\Delta}$ is intersected, $\sigma_{\rm in}=1$ for any
function $\varepsilon(t,k)$. Consequently, the shell initially
moves in the region $R_{+}$. When the radius $\rho_{h}$ is
intersected, $\sigma_{\rm out}=-1$ and, hence, the shell is in the
region $R_{-}$. The Carter-Penrose diagram (global geometry)
corresponding to this case is shown in Figs.~2b-2d (at $k = 0,-1$)
for various equations of state (see also \cite{BerKuzTkach3}).
Below, we will give the Carter-Penrose diagrams only for the case
where $P=\varepsilon/3$, i.~e., $\alpha=1/3$, since the
corresponding diagrams for other equations of state with
$\alpha=const$ are constructed in a similar way. The
Carter-Penrose diagrams for a closed universe will be the same as
those for an open one (see Figs.~2b-2d). For a contracting shell,
the signs of $\sigma$ at which the shell intersects the radii
$\rho_{h}$ and $\rho_{\Delta}$ will remain the same. The
corresponding Carter-Penrose diagram for a closed geometry is
shown in Fig.~2e. Additional peculiarities appear for a closed
universe, i.~e., at $k = 1$. For example, the expansion of a
closed universe changes to its contraction, while we see from the
Carter-Penrose diagram for a closed universe (see Fig.~2b) that
the expansion of the universe cannot change to its contraction,
since there is no region $T_{-}$ for a closed Friedman world. In
particular, this is because a time interval will always be found
when this diagram will not be valid or, more specifically, the
condition $m>m_{3}$ will be violated. We will give an answer to
this question (and to similar questions for other diagrams) at the
end of this section.

\subsection{The Case of $m_{0}<m<m_{3}$}

This case differs from the previous one only in that the radii
$\rho_{h}$ and $\rho_{\Delta}$ are interchanged. The potential and
the Carter-Penrose diagram for an expanding shell for an equation
of state with $\alpha=1/3$ are shown, respectively, in Figs.~2f
and $2g$ ($k = 0,\pm1$). The Carter-Penrose diagrams for a
contracting shell and for other values of $\alpha$ are constructed
in much the same way as in the previous case.

\subsection{The Case of $m_{2}<m<m_{0}$}

The effective potential in this case shown in Fig. 2h has a region
where $U(\rho)>0$. In this case, the shell bounces, i.~e., the
contraction and expansion are interchanged, at $U(\rho)=0$. If the
shell begins its motion from the coordinate origin ($\rho(0)=0$),
then the expansion of the shell changes to its contraction and, in
the long run, it will contract into a singularity. The
Carter-Penrose diagram for a closed world is shown in Fig.~3a. If,
alternatively, the shell begins to contract from infinity, then
this contraction will change to its expansion and the shell will
again expand to infinity. The corresponding Carter-Penrose diagram
for a closed world is shown in Fig.~3b.

\subsection{The Case of $m_{1}<m<m_{2}$}

For a shell contracting from infinity, the situation will not
change compared to the previous case. However, for a shell
expanding from the coordinate origin, the situation will change
radically. Now, $\sigma_{\rm out}=+1$. The corresponding graph of
the potential and the Carter- Penrose diagram for a closed world
are shown in Figs.~3c and 3d.

\subsection{The Case of $m<m_{1}$}

The potential for this case is shown in Fig.~3e. The situation
where the shell expands from the coordinate origin will not change
compared to the previous case, while the situation where the shell
contracts from infinity differs in that the radius $\rho_{1}$ will
be under the graph of the potential (if it will exist at all by
that time). Two alternatives are possible. If the radius
$\rho_{1}$ is absent at the time when the shell intersects the
radius $\rho_{\Delta}$, then the situation is reduced to the
previous case. If, alternatively, the radius $\rho_{1}$ exists at
the time when the shell intersects the radius $\rho_{\Delta}$,
then $\sigma_{\rm in}$, will change its sign or, more
specifically, $\sigma_{\rm in}=-1$. The Carter-Penrose diagram for
this alternative is shown in Fig.~3f.

The classification under consideration allows the dynamics of the
vacuum shell to be completely described without restricting
generality to a short time interval $t$. Indeed, let the mass $m$
be fixed and the condition $m>m_{3}$ be satisfied for some short
time interval. The parameter $m_{3}$ increases with time $t$ and
will become larger than $m$ at some time. The condition
$m_{0}<m<m_{3}$ will then be satisfied and the vacuum shell will
satisfy the corresponding solution for this new inequality
depending on whether it intersected other characteristic radii or
not. The entire subsequent dynamics of the shell can be traced in
a similar way.

The evolution of a contracting vacuum shell can be considered just
as the evolution of an expanding one, since the Friedman Eqs.~(5)
are invariant with respect to the change of sign of the time $t$
to $-t$. Let the vacuum shell begin its motion from the coordinate
origin at $t = 0$. The parameter $m$ will then be larger than all
of the other mass parameters $m_{i}=(m_{0},m_{1},m_{2},m_{3})$,
because $\varepsilon\to\infty$ and $m_{i}\to0$ when $t\to0$.

Let an open or flat universe initially exist inside the vacuum
bubble. Nothing will hinder the expansion of the vacuum shell.
Depending on the relation between the parameters, several
situations can arise. Either the shell will intersect the radius
$\rho_{\Delta}$ and then the radius $\rho_{h}$ or an exchange
between the radii will first take place,
$\rho_{\Delta}\lessgtr\rho_{h}$ (since $m_{3}$ increases linearly
with time, the parameter $m_{3}$ will become larger than a given m
at some time). If $m>m_{0}(t\rightarrow\infty)=4/(27\pi S)$, then
the shell will just go to infinity. If, alternatively,
$m<m_{0}(t\rightarrow\infty)=4/(27\pi S)$, then the inequality
$m<m_{0}$ will hold after some time (i.~e., the potential will
intersect the $U=0$ axis and a bounce point will emerge). However,
since the universe inside the bubble is open, the expansion cannot
change to contraction, i.~e., the vacuum shell can pass only into
the region to the right of the graph of the potential (if the
shell passed into the region to the left of the potential, then it
would bounce at the bounce point and would contract). In the
region to the left of the potential, the vacuum shell would
continue its expansion, going to infinity. There could also be
other special cases during the expansion. For example,
$m<m_{2}(t\rightarrow\infty)$, but this case is similar to the
previous one. Thus, generally, the Carter-Penrose diagram evolves
with time and this evolution is described in different time
intervals by the above diagrams.

If, alternatively, a closed universe exists inside the bubble,
then there always comes a time when
$m<m_{0}(t\rightarrow\infty)=4/(27\pi S)$, since the expansion
should change to contraction. The vacuum shell can then be located
only to the left of the potential and the reflection from the
bounce point is possible (i.~e., the expansion will change to
contraction). In the long run, such a shell will contract
(collapse) into a singularity.

Qualitatively, the embedding diagrams \cite{Zeldov} for vacuum
shells in the Friedman-Schwarzschild world are shown in Fig.~4a
for open and flat Friedman worlds and in Figs.~4b and 4c for a
closed Friedman world (see also \cite{FrolMarMuk,Novik,Novik1}).
We can see from the Carter-Penrose diagram that semiclosed worlds
are formed in almost all cases of shell evolution. This is because
the shell moves in the region $R_{-}$ of the Schwarzschild world,
while the regions $R_{+}$ and $R_{-}$ are connected by a tunnel
(wormhole).

\subsection{APPROXIMATE SOLUTION}

In certain limiting cases, the equation of motion of a vacuum
shell in the Friedman-Schwarzschild world can be solved
approximately. When the shell contracts from infinity and the
effective potential intersects the $U = 0$ axis, the term
\begin{equation}
 -\frac{m}{\rho}\left(1-\frac{\varepsilon}{6\pi
 S^2}\right)-\frac{m^2}{16\pi^2 S^2\rho^4}.
 \label{potential2}
\end{equation}
can be neglected in the effective potential (13). The equation of
motion of the shell (13) will then be significantly simplified and
can be reduced to
\begin{equation}
 \left(\frac{d\rho}{d\tau}\right)^2=\phi^2\rho^2-1,
 \label{eqmot2}
\end{equation}
where we denote $\phi=2\pi S+\varepsilon/(3S)$. In this case,
Eq.~(12) for joining the inner Friedman metric and the metric on
the shell can be rewritten as
\begin{widetext}
\begin{equation}
 \left[1+\left(\frac{d\rho}{d\tau}\right)^2\right]
 \dot\rho^2-2\left(\frac{d\rho}{d\tau}\right)^2H\rho
 \dot\rho+\left(\frac{d\rho}{d\tau}\right)^2
 \left(H^2\rho^2-1\right)=0,
 \label{shifka2}
\end{equation}
\end{widetext}
where $H=\dot a/a=(8\pi\varepsilon/3)^{1/2}$ is the Hubble
constant. From two equations, (\ref{eqmot2}) and (\ref{shifka2}),
we obtain
\begin{equation}
 \dot\rho=(H\sqrt{\phi^2\rho^2-1}
 \pm|\phi-4\pi S|)\frac{\sqrt{\phi^2\rho^2-1}}{\phi^2\rho}\,.
 \label{34}
\end{equation}
In this equation, we can already assume that $\rho$ depends only
on $t$. Integrating this equation, we will obtain the function
$\rho(t)$ and then can find the function $\tau(t)$ using Eq.
(\ref{shifka}) at $\rho=aq$:
\begin{equation}
 \dot\tau^2=1-\left(\dot\rho-H\rho\right)^2.
 \label{dottau}
\end{equation}
Finding the inverse function $t=t(\tau)$ from this equation, we
will ultimately obtain the function  $\rho(\tau)$. For a linear
equation of state ($k = 0$) and taking into account the solution
of the Friedman equations (\ref{frsolution}), we have the relation
$H=\dot{a}/a=n/t$. Let us find an asymptotic solution to
Eq.~(\ref{34}) for $t\rightarrow\infty$. In this limit,
Eq.~(\ref{34}) can be rewritten as
\begin{equation}
 \frac{d\rho}{dt}=\frac{\sqrt{\left(2\pi S\rho\right)^2-1}}
 {\left(2\pi S\rho\right)^2}\left[\frac{n}{t}
 \sqrt{\left(2\pi S\rho\right)^2-1}\pm2\pi S\right].
\end{equation}
Integrating the latter equation at $n\neq1$  yields
\begin{equation}
 \rho=\frac{1}{2\pi S}\sqrt{1+
 \frac{(2\pi S)^2[t-Bn(n-1)t^n]^2}{(n-1)^2}},
\end{equation}
where $B$ is the constant of integration. Accordingly, at $n=1$,
we obtain
\begin{equation}
 \rho=\frac{1}{2\pi S}\sqrt{1+(2\pi S)^2t^2(B+\ln{t})^2}.
\end{equation}
In the limit of $t\rightarrow\infty$ under consideration, we
ultimately obtain
\begin{equation}
 \rho\simeq\left\{
 \begin{array}{lr}
  [t+B(1-n)nt^n]/(1-n), & n\neq1; \\
  t\ln{t}, & n=1.
 \end{array}
 \right.
 \label{rho39}
\end{equation}
It follows from this solution and Eq.~({\ref{dottau}) that
$d\tau/dt=0$, i.~e., the dependence of $t$ on $\tau$ vanishes in
the limit under consideration.

The equation of motion of the shell can also be solved in the
other limiting case where $t\rightarrow0$. In this limit, the
equation of motion of the shell is
\begin{eqnarray}
 \frac{d\rho}{dt}=\frac{n}{t}\rho\pm1.
\end{eqnarray}
The solution to this equation is
\begin{equation}
 \rho\simeq\left\{
 \begin{array}{lr}
  Ct^n\pm t/(1-n), & n\neq1; \\
  Ct\pm t\ln{t}, & n=1,
 \end{array}
 \right.
 \label{rho41}
\end{equation}
where $C$ is the constant of integration. In this limiting case,
there is no dependence of $t$ on $\tau$ either.

In a similar way, we can find an approximate solution to the
equation of motion of the vacuum shell when it moves while being
located to the left of the potential. In this case, terms of the
form
\begin{equation}
 -\frac{\varepsilon^2\rho^2}{9S^2}-4\pi^2 S^2\rho^2-
 \frac{4\pi}{3}\varepsilon\rho^2.
\end{equation}
can be neglected in the potential.

The corresponding solutions to the equation of motion of the
vacuum shell in the limits $t\to\infty$ and $t\rightarrow0$, are
similar in form to (\ref{rho39}) and (\ref{rho41}).

\section{CONCLUSIONS}

We considered the dynamics of a thin vacuum shell in the
Friedman-Schwarzschild world. The total mass m (Schwarzschild
mass) of the shell measured by an observer at spatial infinity is
a convenient parameter for the classification of the possible
types of its evolution. This mass with the gravitational mass
defect includes the total energy of the inner Friedman world and
the total energy of the shell with its surface tension energy and
its kinetic energy. The classification under consideration allows
the dynamics of the vacuum shell to be completely described
without restricting generality to a short time interval. The end
result of the evolution of the vacuum shells under consideration
in the Friedman-Schwarzschild world was shown to be the for motion
of black holes and wormholes with baby universes inside in a wide
range of initial conditions parameterized by the total initial
shell mass. The interior of this world can be a closed, flat, or
open Friedman universe. In the same way, more complex
configurations, for example, where another bubble inside which a
world, other than the Friedman one can be located, is formed
within one bubble \cite{Sato2}, can be investigated using the
method of an effective potential. Such configurations, where the
evolution of the inner and outer bubbles is determined by the
metrics inside and outside the shell, can be analyzed by the
method of an effective potential individually. It should be noted
that the method of an effective potential is inapplicable in the
situation where the bubbles intersect. In the case of very small
bubbles, where the bubble interior is inhomogeneous due to edge
effects and the Friedman equations are inapplicable, we go beyond
the scope of the formalism under consideration.

\begin{widetext}
\begin{figure}[b]
 \begin{center}
  \includegraphics{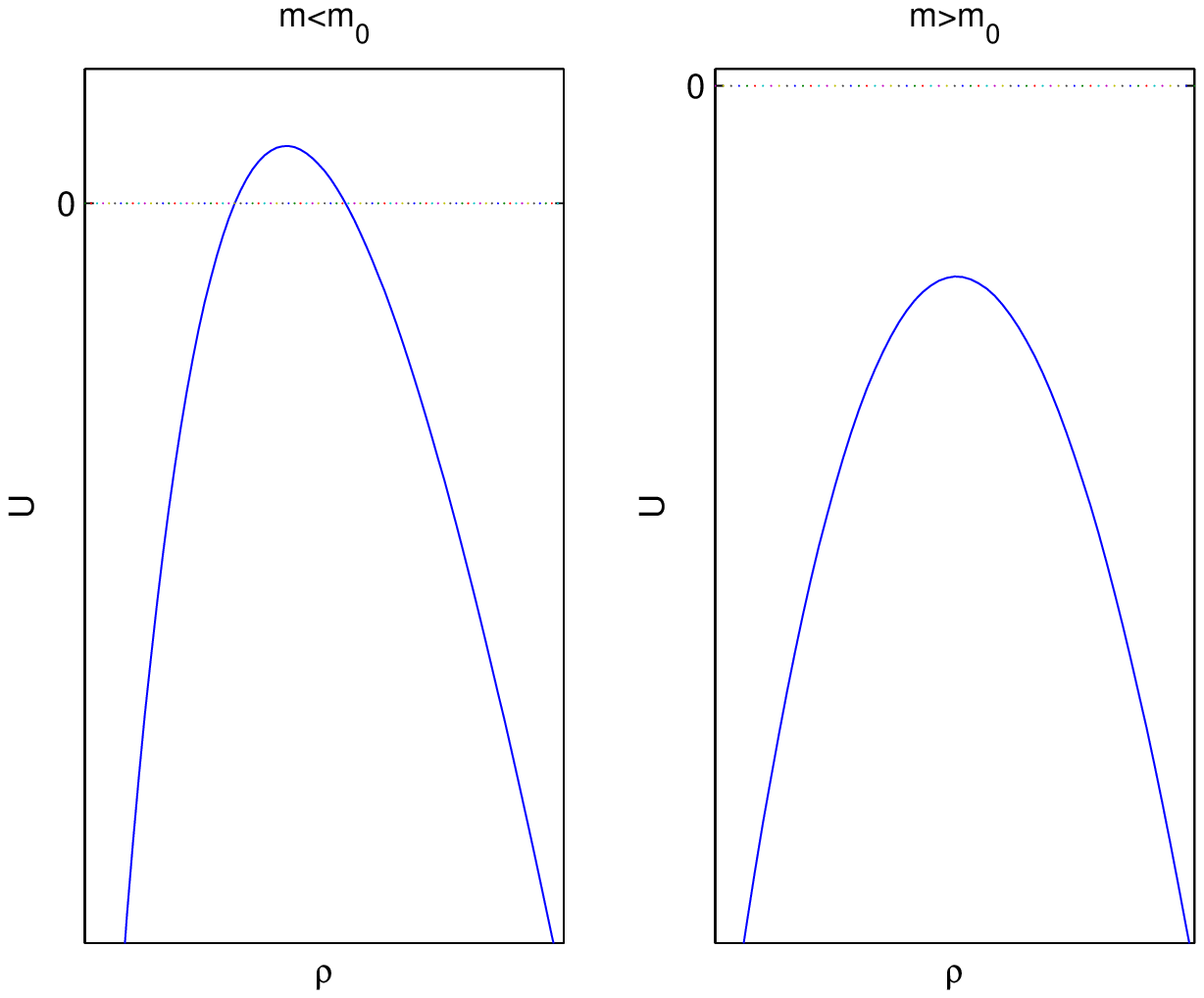}
 \end{center}
 \caption{Effective potential $U(\rho)$ from Eq.~(\ref{potential})
 as a function of the relation between the total mass m of the
 shell and the parameter $m_0$ from Eq.~(\ref{m0}): (a) $m <
 m_{0}$ and (b) $m > m_{0}$.}
 \label{fig1}
\end{figure}
\end{widetext}

\begin{figure}[t]
 \begin{center}
 \includegraphics{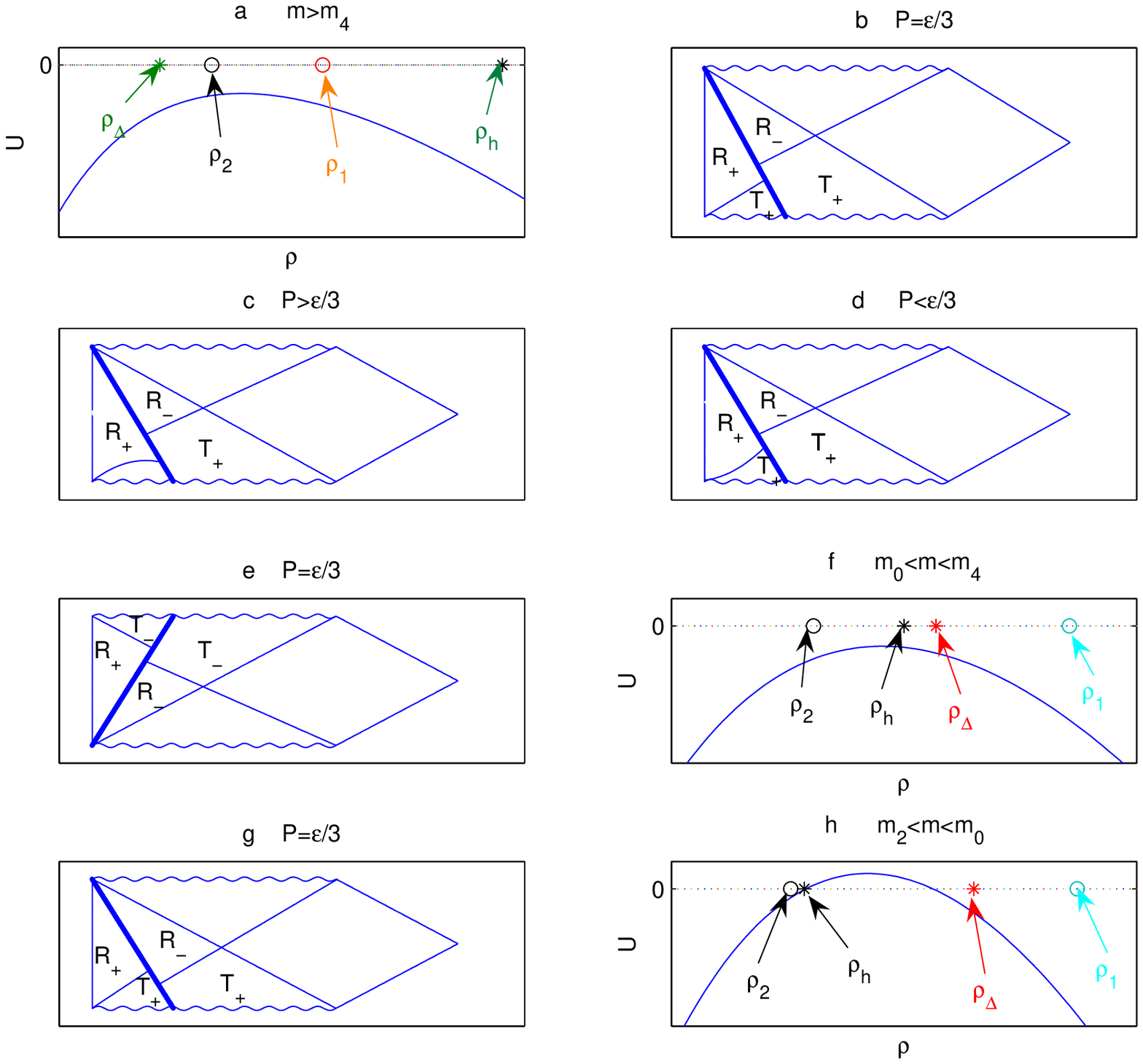}
 \end{center}
 \caption{Effective potentials $U(\rho)$ and Carter-Penrose
 diagrams for the global geometry of a moving vacuum shell in the
 Friedman-Schwarzschild world.}
 \label{fig2}
\end{figure}

\begin{figure}[t]
 \begin{center}
 \includegraphics{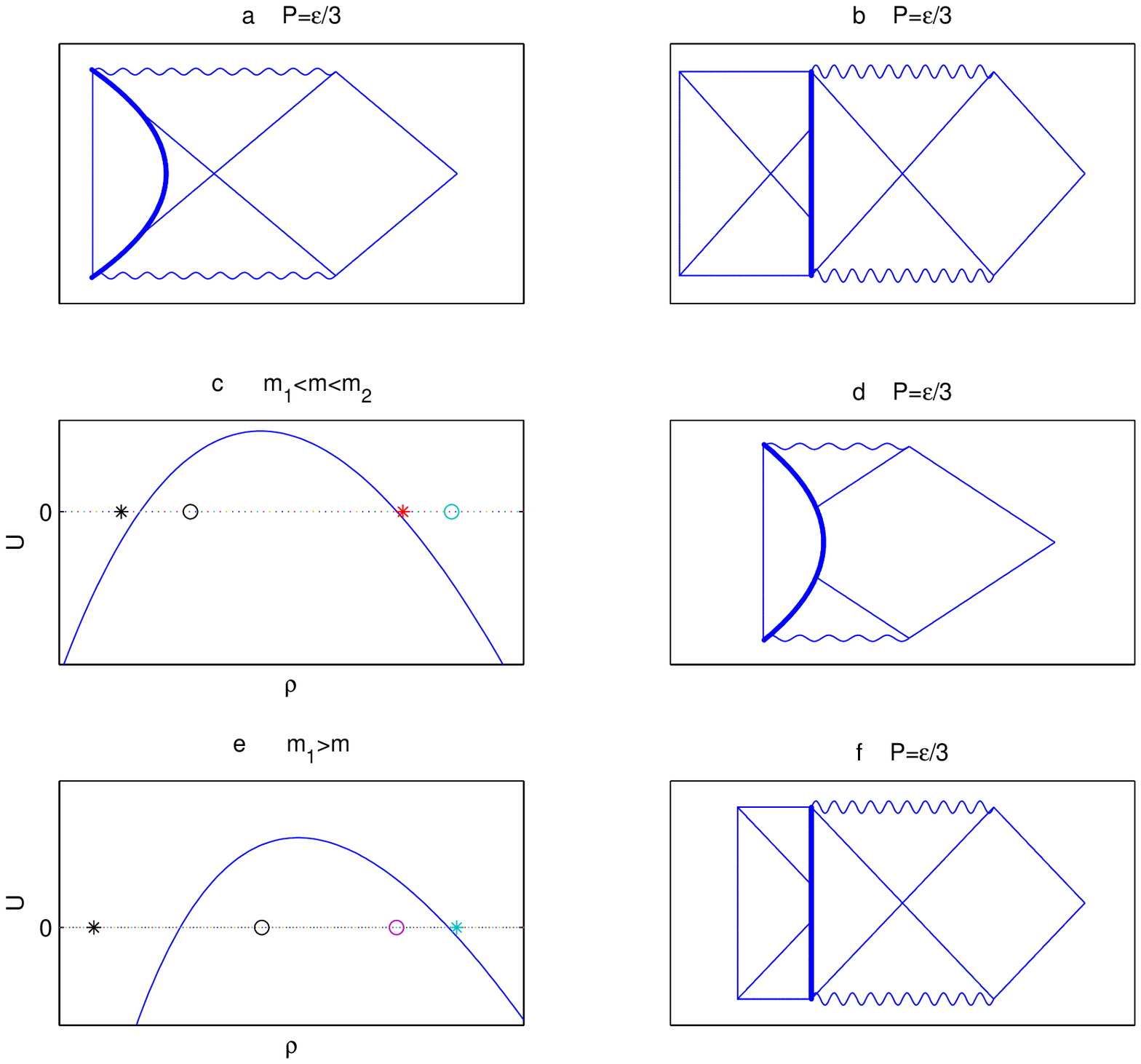}
 \end{center}
 \caption{Effective potentials $U(\rho)$ and Carter-Penrose
 diagrams for the global geometry of a moving vacuum shell in the
 Friedman-Schwarzschild world.}
 \label{fig3}
\end{figure}

\begin{figure}[t]
 \begin{center}
 \includegraphics{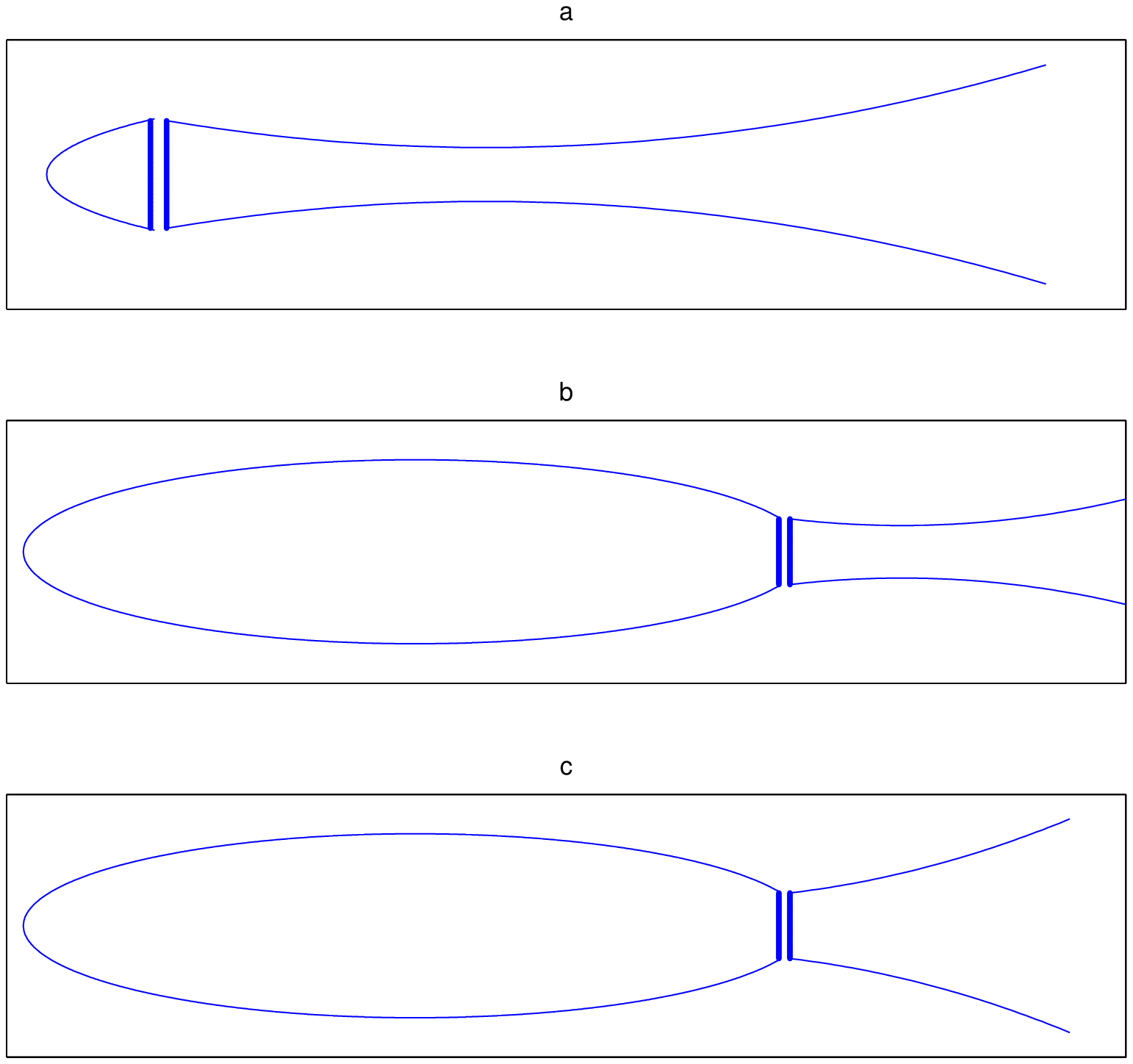}
 \end{center}
 \caption{Embedding diagrams for a shell in the Friedman-Schwarzschild
 world containing open and flat (a) and closed (b,c) Friedman worlds,
 respectively.}
 \label{fig4}
\end{figure}


\begin{thebibliography}{99}
\bibitem{Kirznits} D. A. Kirzhnits, Pis'ma Zh. Eksp. Teor. Fiz. 15
(12), 745 (1972) [JETP Lett. 15 (12), 529 (1972)].

\bibitem{Kirznits2} D. A. Kirzhnits and A. D. Linde, Phys. Lett. B
42, 471 (1972).

\bibitem{ZeldovKobzarOcyn} Ya. B. Zel'dovich, I. Yu. Kobzarev, and
L. B. Okun', Zh. Eksp. Teor. Fiz. 67 (1), 3 (1974) [Sov. Phys.
JETP 40 (1), 1 (1974)].

\bibitem{KobzOkunVoloshin} I. Yu. Kobzarev, L. B. Okun', and M. B.
Voloshin, Yad. Fiz. 20, 1229 (1974) [Sov. J. Nucl. Phys. 20, 644
(1975)].

\bibitem{Col} S. Coleman, Phys. Rev. D: Part. Fields 15, 2929 (1977).

\bibitem{CalCol} C. G. Callan and S. Coleman, Phys. Rev. D: Part.
Fields 16, 1762 (1977).

\bibitem{ColLuc} S. Coleman and F. de Luccia, Phys. Rev. D: Part.
Fields 21, 3305 (1980).

\bibitem{Israel} W. Israel, Nuovo Cimento B 44, 1 (1966).

\bibitem{BerKuzTkach} V. A. Berezin, V. A. Kuzmin, and I. I.
Tkachev, Phys. Rev. D: Part. Fields 36, 2919 (1987).

\bibitem{Sato} K. Sato, M. Sasaki, H. Kodama, and K. Maeda, Prog.
Theor. Phys. 65, 1443 (1981).

\bibitem{SatoSasKodMae1} K. Sato, M. Sasaki, H. Kodama, et al.,
Phys. Lett. B 108, 103 (1982).

\bibitem{Sato1} H. Kodama, M. Sasaki, and K. Sato, Prog. Theor.
Phys. 68, 1979 (1982).

\bibitem{Sato2} H. Kodama, M. Sasaki, K. Sato, et al., Prog.
Theor. Phys. 66, 2052 (1981).

\bibitem{BerKuzTkach1} V. A. Berezin, V. A. Kuz'min, and I. I.
Tkachev, Zh. Eksp. Teor. Fiz. 86 (3), 785 (1984) [Sov. Phys. JETP
59 (3), 459 (1984)].

\bibitem{BerKuzTkach2} V. A. Berezin, V. A. Kuz'min, and I. I.
Tkachev, Pis'ma Zh. Eksp. Teor. Fiz. 41 (10), 446 (1985) [JETP
Lett. 41 (10), 547 (1985)].

\bibitem{BerKuzTkach6} V. A. Berezin, V. A. Kuzmin, and I. I.
Tkachev, Phys. Lett. B 124, 479 (1983).

\bibitem{BerKuzTkach4} V. A. Berezin, V. A. Kuzmin, and I. I.
Tkachev, Phys. Lett. B 120, 91 (1983).

\bibitem{BerKuzTkach5} V. A. Berezin, V. A. Kuzmin, and I. I.
Tkachev, Phys. Lett. B 130, 23 (1983).

\bibitem{IpserSikivie} J. Ipser and P. Sikivie, Phys. Rev. D:
Part. Fields 30, 712 (1984).

\bibitem{Aurilia} A. Aurilia, G. Denardo, F. Legovivni, and E.
Spallucci, Phys. Lett. B 147, 258 (1984).

\bibitem{Aurilia1}  A. Aurilia, G. Denardo, F. Legovivni, and E.
Spallucci, Nucl. Phys. B 252, 523 (1985).

\bibitem{Aurilia2} A. Aurilia, M. Palmer, and E. Spallucci, Phys.
Rev. D: Part. Fields 40, 2511 (1989).

\bibitem{Aguirre} A. Aguirre and M. C. Johnson, Phys. Rev. D:
Part. Fields 72, 103525 (2005).

\bibitem{BlGuenGuth} S. K. Blau, E. I. Guendelman, and A. H. Guth,
Phys. Rev. D: Part. Fields 35, 1747 (1987).

\bibitem{Kardashev} N. S. Kardashev, I. D. Novikov, and A. A.
Shatskiy, Int. J. Mod. Phys. D 16, 909 (2007).

\bibitem{DokCher}. V. I Dokuchaev and S. V. Chernov, Pis'ma Zh.
Eksp. Teor. Fiz. 85 (12), 727 (2007) [JETP Lett. 85 (12), 595
(2007)].

\bibitem{DokCher1} S. V. Chernov and V. I. Dokuchaev, Classical
Quantum Gravity 25, 015004 (2008); arXiv:0709.0616v1 [gr-qc]

\bibitem{Rubakov} V. A. Rubakov, Pis'ma Zh. Eksp. Teor. Fiz. 39
(2), 89 (1984) [JETP Lett. 39 (2), 107 (1984)].

\bibitem{BerKuzTkach3} V. A. Berezin, V. A. Kuz'min, and I. I.
Tkachev, Zh. Eksp. Teor. Fiz. 93 (4), 1159 (1987) [Sov. Phys. JETP
66 (4), 654 (1987)].

\bibitem{IshakLake} M. Ishak and K. Lake, Phys. Rev. D: Part.
Fields 65, 044011 (2002).

\bibitem{FrolMarMuk}Ya. B. Zel'dovich, Zh. Eksp. Teor. Fiz. 43,
1037 (1962) [Sov. Phys. JETP 16, 732 (1962)].

\bibitem{Novik} V. P. Frolov, M. A. Markov, and V. F. Mukhanov,
Phys. Rev. D: Part. Fields 41, 383 (1990).

\bibitem{Novik1} I. D. Novikov, Pis'ma Zh. Eksp. Teor. Fiz. 3 (5),
223 (1966) [JETP Lett. 3 (5), 142 (1966)].

\bibitem{Zeldov} I. D. Novikov, Vestn. Mosk. Univ., Ser. 3: Fiz.,
Astron., No. 5, 90 (1962).

\end{thebibliography}
\end{document}